\documentstyle[a4,12pt,axodraw]{article}

\def\beq{\begin{equation}}
\def\eeq{\end{equation}}
\def\bqa{\begin{eqnarray}}
\def\eqa{\end{eqnarray}}

\def\rw{\rightarrow}

\def\eve{\; ,\;\;\;\;}

\def\d3#1{\frac{d^3 p_#1}{(2\pi)^3 \; 2E_#1}}

\def\Jpsi{J/\psi}
\def\Upsi{\Upsilon}
\def\alfa{\alpha}
\def\alfas{\alpha_{s}}
\def\<{\left\langle}
\def\>{\right\rangle}
\def\gc{\<{\frac{\alpha_s}{\pi}}G^{\mu\nu}G_{\mu\nu}\>}

\begin{document}

\begin{center}
%%%%%%%% TITULO %%%%%%
{\LARGE \bf Dynamical gluon mass corrections in heavy quarkonia
decays}

\vskip 1.0cm
                 
{\large A.\ Mihara~\footnotemark 
\footnotetext{e-mail : mihara@ift.unesp.br} and 
A.\ A.\ Natale~\footnotemark
\footnotetext{e-mail : natale@ift.unesp.br}}\\[.2cm]
 
{\it Instituto de F\'{\i}sica Te\'orica \\
Universidade Estadual Paulista \\
Rua Pamplona, 145 \\
01405-900, S\~ao Paulo - SP \\
Brazil}

\vskip 1.0cm

\end{center}

{\bf Abstract}
Using the expression of the
dynamical gluon mass obtained through the operator product expansion
we discuss the relevance of gluon mass effects in the decays
$V \rw hadrons$ ($V = \Jpsi ,\;\Upsi$).
Relativistic and radiative corrections are also introduced to calculate
$\alfas(m_c)$ and $\alfas(m_b)$ comparing them
with other values available in the literature. 
The effects of dynamical gluon masses are negligible for $\Upsi$ decay 
but important for $\Jpsi$ decay.

%\vskip 1.5cm
%\pacs
{PACS number(s): 13.20.Gd, 13.25.Gv, 14.40.Gx}

\newpage  

Since the original suggestion by Appelquist and Politzer~\cite{AP},
the annihilation of heavy quark pairs into gluons or gluons plus
a photon has been recognized as an apparently excellent process
for testing the basic ideas of QCD and for measuring 
the strong coupling $\alfas$ with high precision.

One can determine $\alfas$ at low energies by comparing the 
experimental $\Upsi$ and $\Jpsi$  branching ratios
\beq
R_V = \frac{\Gamma(V\rw ggg)}{\Gamma(V\rw ee)}\eve V=\Upsi,\;\Jpsi.
\eeq
with the theoretical prediction. This is given, at the lowest order in
perturbation theory, by
\beq
R_V = \frac{10\;(\pi^2 -9)}{81\pi e_{q}^2\;\alfa(M_V)}\;\alfas^3 \; ,
\label{eq:rvlo}
\eeq
where $e_{q}$ is the quark charge, $\alfa$ is the electromagnetic coupling and
$M_V$ is the quarkonium mass. In addition, the experimental data for the
radiative process $V \rw \gamma \, + \, X$ is an important source of
information about the QCD dynamics, which predicts a photon spectrum nearly
linear in $z=2 E_{\gamma}/M_V $, where $E_{\gamma}$ is the photon energy.

The match of the experimental and theoretical 
determinations of $\alfas$ and the photon spectrum with high accuracy
has originated a long series of discussions since the pioneer work of
Ref.~\cite{AP}. Corrections due to relativistic and radiative QCD effects
were shown to modify Eq.(\ref{eq:rvlo}). Other effects, as the existence of an
effective gluon mass or beyond standard model contributions, are also
possible explanations for the claimed differences between theory and
experimental data. In this work we introduce the concept of a dynamical
gluon mass and verify how it affects quarkonia decays. The fact that
we consider a momentum-dependent gluon mass clearly modifies
previous discussions on this problem.
 
Parisi and Petronzio \cite{PP} were the first to discuss
some discrepancies between QCD results and the ones obtained from $\Jpsi$
decays. They introduced an effective gluon mass to improve the comparison
between experiment and theory in the case of $\Jpsi$ decays. This is not the
unique alternative used to solve the problems in this context but this line of
thought does give reasonable results, and the possible effects of a gluon
mass in quarkonia decays were considered several times in the
literature (see \cite{CF1}-\cite{AM} and references therein).
In Refs.~\cite{PP,CF1,CF2} the value of the gluon mass was determined
calculating the CM energy spectrum of direct photons produced by 
$\Jpsi$ and $\Upsi$ decays ($\Jpsi \, , \Upsi \rw gg\gamma$). Afterwards the 
spectra were compared with experimental data, which clearly favoured the 
case of a massive gluon and fixed its mass. It is interesting to note that
the explanation of the photon spectrum in the $\Jpsi$ decay demanded a gluon
mass $M_g \simeq 0.66$ GeV, whereas in the $\Upsi$ decay the most suitable mass
value raised to $M_g \simeq 1.2$ GeV~\cite{CF1,CF2}. There are some points
that may be criticized in these results. With the knowledge accumulated in the 
past decades about the theory of the strong interactions it is
obvious that we cannot have a bare gluon mass in QCD. 
In Refs.~\cite{CF1,CF2} the
gluon masses differ in the $\Jpsi$ and $\Upsi$ decays in an unnatural way: as
we increase the quarkonium mass the gluon mass needed to
explain the photon spectrum is also increased. Finally, the explanation of
the bending of the photon spectrum at large $z$ in these decays may be 
due to the radiation of additional gluons as described by Field~\cite{F}. 

It is clear that QCD does not admit a bare gluon mass.
However, this theory
may have a dynamical gluon mass~\cite{C} and recent simulations of QCD on
the lattice strongly support this possibility~\cite{Lat}. Therefore, the
proposal of Parisi and Petronzio~\cite{PP} may indeed be realized in Nature,
but in a more subtle way. This will be our working hypothesis. The next
question is how to consider this dynamical mass in quarkonium decays.
The introduction of a gluon mass scale in the calculation of some hadronic
processes has been performed in a rather heuristic way (see, for instance,
Ref.~\cite{HKN}). Only recently a more formal handling of gluon masses
has been discussed~\cite{FPP}. In principle, we can justify the approach
of Ref.~\cite{HKN} within the dynamical perturbation theory proposed
by Pagels and Stokar many years ago~\cite{PS}, which basically imply
in the use of the running gluon mass in the gluon propagator. This is the
approach that we will follow here.

In order to  obtain the asymptotic behavior of the gluon mass we
can rely on the operator product expansion (OPE). This asymptotic behavior
was obtained in Ref.~\cite{L}: 
\beq 
M_g^2 (P^2) \sim  \frac{34 N \pi^2}{9(N^2-1)} 
\frac{\gc}{P^2}, 
\label{mgope} 
\eeq 
where $P^2 (\equiv -p^2 )$ is the gluon momentum in Euclidean space, 
$\gc$ is the gluon condensate~\cite{SVZ}, and $N = 3$ for QCD. At this
point we see the difference with previous work~\cite{CF1,CF2}: the dynamical
mass is connected with the gluon condensate and it decreases with
energy. Therefore, we can expect different results from the ones of
Refs.~\cite{CF1} and ~\cite{CF2}. 

Equation (\ref{mgope}) gives the asymptotic behavior of the gluon mass
but we also need its expression in the infrared region. If we define
\beq 
m_g^2 \equiv \left( \frac{34 N \pi^2}{9(N^2-1)} \gc \right)^{1/2}, 
\label{mg}
\eeq 
it was recently verified that the value of the dynamical gluon mass
in the infrared is well described by the OPE value frozen at the scale 
$m_{g}$~\cite{GN}, {\sl i.e.} we can write
\beq 
M^2_g (P^2)= m^2_g \theta (m^2_g-P^2) + \frac{m^4_g}{P^2} 
\theta (P^2-m^2_g). 
\label{mgrun} 
\eeq 
This is the expression which will be used in the calculation of
quarkonium decays.

As discussed in Ref.\cite{PP}, the gluon mass implies that the
branching ratio $R_V$ must be changed to $R_V \cdot f_3(\eta)$, where
\beq
f_3(\eta) = \frac{\Gamma ( V \rw ggg )|_{m_{g}}}
{\Gamma ( V \rw ggg)|_{m_{g}=0}},
\label{f3}
\eeq
is a  function of $\eta\equiv 2m_{g}/M_{V}$. 
This simple factorization is a consequence
of the fact that most of the gluon mass contribution in the calculation of
$\Gamma ( V \rw ggg)$ comes from the phase space integration~\cite{PP,MP}.
Therefore, we proceed in the same way as in \cite{PP} to compute $f_3$, with
the difference that the mass is now a function of each final state gluon
momentum.

In order to compute the function $f_3$ as a function of $\eta$ we 
made use of the package COMPHEP~\cite{BO}.
Initially we considered the case where a bare gluon mass was introduced in the
matrix element and in the phase space. The result is given by the solid line in
Fig.1, which is identical to the one of Ref.\cite{PP}. We then introduced a
bare mass only in the phase space integration, obtaining the thin dashed line
of Fig.1 (compatible with Ref.\cite{CF1}). This result shows that the mass
effect can be accounted for within this simple approximation. Once we verified
that this approximation reproduces most of the mass effect, the
calculation to obtain $f_3$ was performed by Monte Carlo integration
considering the dynamical gluon mass contribution (Eq.\ (\ref{mgrun})) only in
the phase space. The result is the thick dashed line depicted in Fig.1.

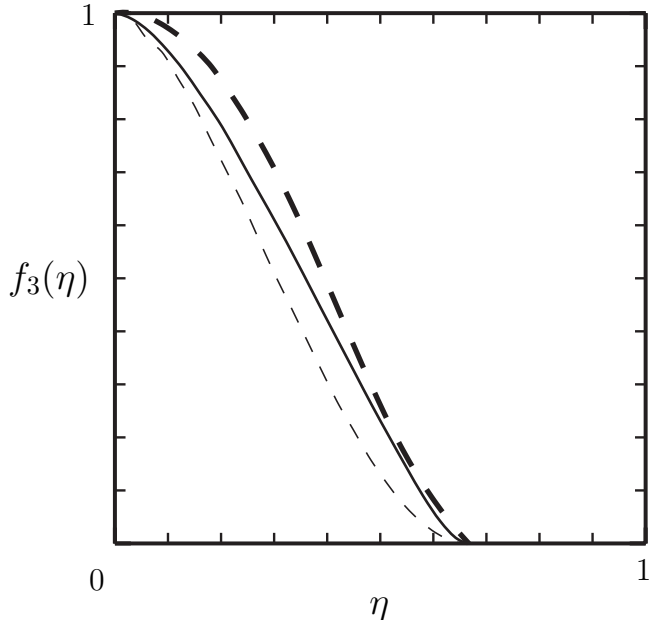
\begin{figure}[ht]
\begin{center}

\begin{picture}(200,200)(0,0)

\LinAxis(0,0)(200,0)(1,10,5,0,1.5)
\LinAxis(0,200)(200,200)(1,10,-5,0,1.5)
\LinAxis(0,0)(0,200)(1,10,-5,0,1.5)
\LinAxis(200,0)(200,200)(1,10,5,0,1.5)

\SetScale{200.} \SetWidth{0.005}
\Curve{(0.000,1.000)(0.032,0.989)
(0.065,0.965)(0.097,0.932)
(0.129,0.893)(0.161,0.847)
(0.195,0.797)(0.257,0.687)
(0.323,0.569)(0.387,0.446)
(0.452,0.322)(0.484,0.261)
(0.517,0.201)(0.549,0.144)
(0.581,0.089)(0.613,0.042)
(0.646,0.008)(0.667,0.00000001031)}

\SetScale{200.} \SetWidth{0.003}
\DashCurve{( 0.000,  1.000)( 0.030,  0.991)( 0.060,  0.952)( 0.090,  0.925)
( 0.120,  0.875)( 0.150,  0.825)( 0.180,  0.763)( 0.210,  0.703)
( 0.240,  0.643)( 0.270,  0.576)( 0.300,  0.510)( 0.330,  0.449)
( 0.360,  0.387)( 0.390,  0.325)( 0.420,  0.265)( 0.450,  0.213)
( 0.480,  0.161)( 0.510,  0.117)( 0.540,  0.078)( 0.570,  0.047)
( 0.600,  0.023)( 0.630,  0.007)( 0.660,  0.000)}{0.03}

\SetScale{200.} \SetWidth{0.01}
\DashCurve{( 0.000000 , 1.000000 )
( 0.030000 , 1.000871 )
( 0.060000 , 0.993082 )
( 0.090000 , 0.978111 )
( 0.120000 , 0.957650 )
( 0.150000 , 0.932619 )
( 0.180000 , 0.901403 )
( 0.210000 , 0.859323 )
( 0.240000 , 0.813942 )
( 0.270000 , 0.762725 ) 
( 0.300000 , 0.707710 )
( 0.330000 , 0.646868 )
( 0.360000 , 0.582017 )
( 0.390000 , 0.513134 )
( 0.420000 , 0.443130 )
( 0.450000 , 0.373264 )
( 0.480000 , 0.305006 )
( 0.510000 , 0.240061 )
( 0.540000 , 0.180857 )
( 0.570000 , 0.128632 )
( 0.600000 , 0.082910 )
( 0.630000 , 0.042540 )
(0.667, 0.000) }{0.05}

\Text(0,-10)[rt]{0 } \Text(200,-10)[]{1}
\Text(-10,200)[]{1}
\Text(-25,100)[]{\bf\large$f_{3}(\eta)$}
\Text(100,-25)[]{\large$\eta$}
\end{picture}

\end{center}
{\small\caption{Function $f_{3}(\eta)$ representing the gluon mass corrections.
The solid line represents these corrections when the effect of a bare gluon     
mass is taken into account both in the matrix element and phase space. The 
thin dashed line is obtained considering the effect of a bare gluon mass
only in the phase space. The thick dashed line is our result obtained by
taking into account a dynamical mass as given by Eq.\ (\ref{mgrun}).}}
\label{fig:f3} 
\end{figure}
\vskip 0.1cm

As discussed before, in Refs.~\cite{PP} and \cite{CF1}, the value of the gluon
mass was obtained by calculating the CM energy spectrum of direct photons
produced by $\Jpsi$ and $\Upsi$ decays ($\Jpsi \, , \Upsi \rw gg\gamma$), 
yielding very different values in each case. With the gluon mass obtained
from the fit of the  radiative quarkonia decays, the value of $f_3$ was
determined from a curve  similar to Fig.1, leading to a new determination
of $\alfas$.  Here the gluon mass $m_g$ is given by Eq. (\ref{mg}). If we
consider the gluon condensate value given in Ref.~\cite{SVZ,N}, and the
gluon mass obtained in Refs.~\cite{C,HKN,MNR}, we can
assume $m_g\approx (0.64\pm 0.20)$GeV, where the uncertainty is large
enough to express the various limitations present in the dynamical gluon mass
estimates. It is interesting to see that the central value is of the order
of the value determined by Consoli and Field~\cite{CF1,CF2} to explain
the $\Jpsi \, \rw  \, X + \gamma$ decay ($m_g\approx 0.66$ GeV). With
this value for the gluon mass we certainly do not explain the direct
photon spectrum in the inclusive decay of the $\Upsi$.
However, the explanation of
this process may depend on a complex relationship between
radiation of additional gluons as described by Field~\cite{F} and the
existence of a dynamical gluon mass.

As one can notice in Fig.1, the new curve (the thick dashed one)
is flatter than the curves of Refs.~\cite{PP, CF1} for small values of
$\eta$. This is not so surprising since as $m_g$ decreases $f_3$ tends
to 1. But this feature is strengthened in the case of a dynamical mass
whose momentum dependence is given by Eq.(\ref{mgrun}), because for
small $\eta$ (which means small gluon mass or large quarkonium mass)
the integrated phase space is larger than in the case of a bare gluon
mass. This is so because we have configurations where at least one of the
gluons resulting from the quarkonium decay has its mass effect sharply cut-off
by a large momentum.

Using the values of $f_3$ from Fig.1 we can now calculate $\alfas(m_c)$ and
$\alfas(m_b)$ . Assuming $m_g = (0.64\pm 0.20)$ GeV, we  have ($\eta=2m_g/M_V$)
\beq 
\eta^{(\Jpsi)}=(0.41\pm 0.13) ,
\eeq 
and 
\beq 
\eta^{(\Upsi)}=(0.14\pm 0.04) ,
\eeq
and from the curve of Fig.1, we obtain
$f_{3}(\eta)$, {\it i.e.} the factors due to gluon mass corrections are
\beq
f_{3}(\eta^{(\Jpsi)})=0.47\pm 0.30 \eve f_{3}(\eta^{(\Upsi)})=0.94\pm 0.03 .
\eeq
Our value of $f_{3}(\eta^{(\Jpsi)})$ is consistent with the one in
Ref.\cite{CF1,CF2}, but $f_{3}(\eta^{(\Upsi)})$ is approximately
$30\%$ larger and almost identical to $1$. If the mass
was also introduced in the matrix element, the values of $f_3$ would
be increased by a few percent and no signal of the gluon mass would appear for
$f_{3}(\eta^{(\Upsi)})$.

We can now compute $\alfas$ from the $\Jpsi$ and $\Upsi$ decays, but to do
so we have to take also into account the relativistic and QCD corrections.   
The QCD corrections are introduced through the factor (see \cite{KMRR})
\beq
\left(1+\frac{\alpha_s}{\pi}b_V \right)
\eeq
where $b_V = 1.6,\; 0.43$ for $\Jpsi$ and $\Upsi$, respectively.

Equation (\ref{eq:rvlo}) is obtained assuming that the $q\bar{q}$ pair
annihilation occurs at a point. Relativistic corrections arise if we
consider that this process occurs over a finite volume of radius
$\simeq 1/m_{q}$. Several papers have dealt with this issue. We follow
Ref.~\cite{CHP} and use their  correction factor:
$\gamma^{(c)}=(0.31\pm 0.03)$ for the $\Jpsi$ and $\gamma^{(b)}=
(0.69\pm 0.07)$ for the $\Upsi$.

Equation (\ref{eq:rvlo}) rewritten with the correction factors
mentioned above becomes
\beq
R_V = f_{3}(\eta)\;\left(1+\frac{\alpha_s}{\pi}b_V \right)\;\gamma\;
\frac{10\;(\pi^2 -9)}{81\pi e_{q}^2\;\alfa(M_V)}\;\alfas^3.
\eeq
We then use the experimental values of $R_V$ from \cite{PDG} and \cite{CF2}   
to calculate $\alfas(m_c)$ and $\alfas(m_b)$: 
$R_{\Jpsi}=10.1\pm 0.9$ and
$R_{\Upsi}=32.6\pm 0.8$.
Our results are
\beq
\alfas(m_c)=0.35\pm 0.07 ,
\label{a1}
\eeq
and
\beq
\alfas(m_b)=0.206\pm 0.008.
\label{a2}
\eeq
The values of Eqs.(\ref{a1}) and (\ref{a2}) are compatible with the ones
of Refs.~\cite{CF2,CHP}. Note that Ref.~\cite{CF2} does not consider the
relativistic corrections of Ref.~\cite{CHP}. On the other hand, our result
for $f_{3}(\eta^{(\Upsi)})$  is about $1.6$ times larger than the one of
Refs.\cite{CF1,CF2}. We find that the
$\alfas(m_b)$ determination through $\Upsi$ decays is barely affected by the
dynamical gluon mass effects and, considering the
relativistic corrections, it is totally consistent with the one of
Ref.\cite{CF1} (where the relativistic corrections were not considered).
Contrary to the values of $\alfas(m_c)$ obtained in the references quoted
above our result has a large error, and most of it, as we discuss next,
is due to the uncertainty in the gluon mass.

Assuming that the gluon propagator has an infrared mass scale~\cite{Lat}, we
see several limitations in order to determine $\alfas$ with high
confidence level through heavy quarkonium decays.
The main problem of this approach is that the dynamical gluon mass scale is
poorly known. The several determinations of this mass scale are characterized
by a large range of possible values~\cite{C,HKN,MNR}. Even if we consider
that this mass follows the behavior predicted by OPE~\cite{L}, we know
that the gluon condensate in opposition to the fermionic one, is known with a
large uncertainty~\cite{N}. The extrapolation of this value to the infrared
region also involves a series of approximations~\cite{GN}. Finally, if we
attempt to verify the consistency between measurements of $\alfas$ at different
mass scales, we also may be in trouble if we do not take into account the
gluon mass effect in the running coupling constant~\cite{PC}.

If the gluon propagator has a dynamical mass scale the
strong running coupling constant is modified, at low energy, to~\cite{C,PC}
\beq 
\alpha_s (Q^2) \simeq \frac{12 \pi}{(11N -2n_f)\ln[(Q^2 + \xi
m_g^2)/\Lambda^2]},
\label{npalpha}
\eeq 
where $n_f$ is the number of
fermion flavors, and $\xi \approx 4$ (or even larger~\cite{PC}) is a
parameter determined in Ref.~\cite{C}. This behavior of
the running coupling constant with respect to the gluon mass appears when
solving the Schwinger-Dyson equations for the gluon propagator and the
trilinear vertex. The determination of Eq.(\ref{npalpha})
and its higher order corrections is a
much more complex problem than the determination of the perturbative
expression of $\alpha_s (Q^2)$, and, unfortunately, is not under complete
control as the perturbative one. Therefore, at the scale $m_b$ and up we do not
expect any substantial difference in the phenomenology if we use the
perturbative running coupling constant when checking the consistency of
$\alpha_s (m_b^2)$, obtained through  Eq.(\ref{eq:rvlo}), with others
measurements of $\alpha_s$. At the scale $m_c$ this is not true anymore. In
this case Eq.(\ref{npalpha}) and its higher order corrections (not yet
computed) have to be used. At leading order of both coupling constants
(perturbative and nonpertubative), the difference at the scale $m_c$ in using
one or another expression amounts a factor roughly given by $\ln{(m_c^2 +\xi
m_g^2)}/\ln{m_c^2} \approx 1.6$. This number is large enough to interfere in
high precision measurements of the running coupling constant using $\Jpsi$
decays, and a better understanding of the gluon mass dependence of the running
coupling constant is needed.

In conclusion, we have presented an analysis similar to
that of  Refs.~\cite{PP,CF1,CF2}. Notwithstanding, we considered
the effects of a dynamical gluon mass, as given by Eq.\ (\ref{mgrun}), which
is compatible with the asymptotic behavior predicted 
by OPE ~\cite{L}. Monte Carlo integration of the phase space in quarkonium
decays provided a new curve for the parameter $f_3(\eta)$, which is flatter for
small values of $\eta$ than the previous ones, and reflects the  momentum
dependence of the dynamical gluon mass.
Assuming $m_g\approx (0.64\pm 0.20)$GeV, we have obtained new values for
$f_3(\eta^{\Jpsi})$ and $f_3(\eta^{\Upsi})$.
With these values we calculated $\alfas(m_c)$ and $\alfas(m_b)$ including 
QCD as well as relativistic corrections. 
The values of $\alfas$ which we have obtained, within the assumed error bars
for the gluon masses, are
compatible with the ones in Refs.~\cite{CF2,CHP}. Contrarily to previous
analysis we verified that $\alfas(m_b)$ is not affected by the existence of a
dynamical gluon mass. More rigorous determinations of the gluon mass scale are
necessary in order to determine $\alfas(m_c)$ with high precision by using
heavy quarkonium decays. 

\section*{Acknowledgments}

We are grateful to R.\ Rosenfeld for useful discussions.
This research was supported in part by 
the Conselho Nacional de Desenvolvimento Cient\'{\i}fico e Tecnol\'ogico (CNPq)
(AAN), Coordenadoria de Aperfei\c coamento de Pessoal de Ensino Superior
(CAPES) (AM), Funda\c c\~ao de Amparo \`a Pesquisa do
Estado de S\~ao Paulo (FAPESP) (AAN), and by Programa de Apoio a
N\'ucleos de Excel\^encia (PRONEX).

\end{document}